\documentclass[journal]{IEEEtran}
\usepackage{graphicx}
\usepackage[cmex10]{amsmath}
\usepackage{amssymb}
\usepackage{algorithmic}
\usepackage{cite}
\usepackage{epstopdf}
\usepackage[tight,footnotesize]{subfigure}
\usepackage{url}
\begin{document}

\title{A Game Theory Interpretation for Multiple Access in Cognitive Radio Networks with Random Number of Secondary Users}

%\author{\IEEEauthorblockN{Oscar Filio Rodriguez, Serguei Primak, Abdallah Shami}
%\IEEEauthorblockA{Department of Electrical and \\Computer Engineering\\
%The University of Western Ontario\\
%London, Ontario, Canada\\
%Email: \{ofilioro, slprimak, ashami\}@uwo.ca}
%\and
%\IEEEauthorblockN{Valeri Kontorovich}
%\IEEEauthorblockA{CINVESTAV\\
%Mexico, DF, Mexico\\
%Email: valeri@cinvestav.mx}
%}

\author{
Oscar~Filio~Rodriguez,~\IEEEmembership{Student Member,~IEEE,}
Serguei~Primak,~\IEEEmembership{Member,~IEEE,}% <-this % stops a space
        Valeri~Kontorovich,~\IEEEmembership{Fellow,~IEEE,}
Abdallah~Shami,~\IEEEmembership{Member,~IEEE}
\thanks{O. Filio, S. Primak and A. Shami are with the Bell Centre for Information Engineering (BCIE), Department
of Electrical and Computer Engineering, The University of Western Ontario, London,
ON, N6A 5B9, Canada, e-mail: ofilioro@uwo.ca, slprimak@uwo.ca, ashami@eng.uwo.ca  }
\thanks{V. Kontorovich is with CINVESTAV - Telecommunications Av. Instituto Polit\'ecnico Nacional, 2508 Col. San Pedro Zacatenco C.P. 07360, Mexico City 07360, Mexico, e-mail: valeri@cinvestav.mx.}}%

\markboth{}%
{O. Filio \MakeLowercase{\textit{et al.}}:}

\maketitle

\begin{abstract}
In this paper a new multiple access algorithm for cognitive radio networks based on game theory is presented. We address the problem of a multiple access system where the number of users and their types are unknown. In order to do this, the framework is modelled as a non-cooperative Poisson game in which all the players are unaware of the total number of devices participating (population uncertainty). We propose a scheme where failed attempts to transmit (collisions) are penalized. In terms of this, we calculate the optimum penalization in mixed strategies. The proposed scheme conveys to a Nash equilibrium where a maximum in the possible throughput is achieved.
\end{abstract}

\IEEEpeerreviewmaketitle

\section{Introduction}\label{Sec:Introduction}
Cognitive Radio (CR) has recently emerged  as a promising technology capable of taking advantage of wasted or temporary unused radio spectrum bands and providing high throughput to unlicensed users who transmit within such bands. In this new communication paradigm, two types of users are considered: Primary Users (PU) who are allowed to access the channel at any moment, and Secondary or Cognitive Users (SU). SUs can only access the channel under certain circumstances (\emph{e.g}. when the PU is not present) or under specified power restrictions, such that the quality of service (QoS) of the PU remains unaltered. It is well known that the surrounding radio environment changes due to the random nature of wireless channels, the dynamic topology and the mobility of users. The SUs should be capable of observing such changes, learning and taking decisions accordingly in order to maximize their throughput. Within this context, SUs can be considered as players competing for a specific license band, resulting in different kinds of payoffs for them. Game theory is a very powerful mathematical tool to model these kind of situations where intelligent rational decision-makers are competing for the same resource. This has been well studied and applied to communication systems in several scenarios\cite{BibRef:MackKenzieBook}\cite{BibRef:ZhangBook}. Due to the inherently competitive nature of CR networks, game theory arises as a straightforward approach to deal with several application problems where the SUs can be modelled as the players in conflict (see for instance \cite{BibRef:BeibeiWang10}-\nocite{BibRef:Feng13} \nocite{bibRef:Gharehshiran13}\nocite{BibRef:Zhang12}\!\!\!\!\!\cite{BibRef:Kasbekar12} ). Specifically, the Medium Access Control (MAC) problem has been analyzed using game theory tools in previous literature such as \cite{BibRef:MackKenzieBook}\cite{BibRef:ZhangBook}\cite{BibRef:MacKenzie03infocom}\cite{BibRef:Xiao05}.
 In \cite{BibRef:Xiao05} a game theory model is presented as a starting point for the Distributed Coordination Function (DCF) mechanism in IEEE 802.11.  In the DCF there are no base stations or access points which control the access to the channel, hence all nodes transmit their data frames in a competitive manner. However, as pointed out in \cite{BibRef:Xiao05}, the proposed DCF game does not consider how the different types of traffic can affect the sum throughput of the system. Moreover, in \cite{BibRef:Bianchi00} it is assumed that the total number of players is known in order to evaluate the performance of the DCF. On the other hand, in \cite{BibRef:MacKenzie03infocom} a game-theoretic model of multipacket slotted ALOHA with perfect information is studied. The authors show that in this model, the Nash equilibrium must exist and its stability region is characterized. Furthermore, a pricing strategy based on slotted ALOHA with multipacket reception is proposed in \cite{BibRef:DandanWang06} in order to enforce fairness among the players. In \cite{BibRef:Bae13} the author calculates an optimal access probability based on slotted ALOHA which maximizes the successful delivery probability in CR networks. Nonetheless, the author assumes that the number of transmitters and receivers is always known during the analysis and all users share a common access probability (\emph{i.e.} all users are treated equally.) In \cite{BibRef:Park11} a distributed MAC algorithm with one-slot memory is proposed in order to coordinate the access among the the secondary users and restrict interference to the PU. An optimal probability of attempting to access the channel for the SUs in order to maximize the throughput is obtained. A $p$-persistent protocol to control the selection of the contention window in the IEEE 802.11 backoff algorithm is described in \cite{BibRef:Cali00}. The authors had shown how to maximize the throughput of the scheme. Notwithstanding, the authors in \cite{BibRef:Park11} and \cite{BibRef:Cali00} do not consider either the possibility of having different types of users, nor the randomness in the number of SUs in the system \cite{BibRef:Kwak05}\cite{BibRef:Zheng06lcn}\cite{BibRef:Ghez88}. In contrast, this paper addresses the problem similar to that approached by the Enhanced DCF included in IEEE 802.11e\cite{BibRef:Choi03icc}, which is a natural extension of the DCF mechanism. To be more specific, we provide a different interpretation and analysis in order to solve the problem of multiple access for a heterogenous and random population of SUs, based on Myerson's results for Poisson games\cite{BibRef:Milchtaich03}. This branch of game theory analysis, addresses the problem of games with an uncertain population, in other words, when the number of players\footnote{Throughout this paper the terms players and secondary users shall be used interchangeably.} is unknown and can be modelled as a random variable.

 The paper is organized as follows. Section \ref{sec:PoissonGames} summarizes the theoretical basis of Poisson Games and two multiple access examples are given. Section \ref{sec:GameModel} presents our novel Poisson game model. We calculate the optimal mixed strategies and the optimal penalizations used in the game. Section \ref{sec:PoissonTwoTypesOfPlayers} extends the aforementioned analysis to the case of two types of SUs and provides an accurate analytical approximation to the Pareto frontier. In Section \ref{sec:GameModelwithPUactivity} the impact on the PU based on its activity is considered and the optimal mixed strategies are calculated accordingly. Finally some conclusions are drawn in Section \ref{sec:Conclusion}.

\section{Poisson Games}\label{sec:PoissonGames}
In \cite{BibRef:Milchtaich03} it is established the concept of a Poisson Game as a special case of a more general type of games called Random Player Games. In games with population uncertainty \cite{BibRef:Myerson98}\cite{BibRef:Myerson00}, there is a nonempty finite set of players types $\mathcal{T}$ which is known \emph{apriori}. In the context of communications systems, this set could contain the different types of services offered by the network (voice, video, data, etc).
There is also a finite set of available choices or pure actions $\mathcal{C}$ that a player may take. For instance, these could be all the different transmission powers that the secondary user may utilize\cite{BibRef:Koskie05} or, in the context of this paper, the decision to transmit or not. The set of possible actions is the same regardless of the type of player. The main characteristic of a Poisson Game is that the total number of players of certain types, are modelled as random variables. In this paper we use the definition given by Myerson\cite{BibRef:Myerson98} which presents a Poisson Game $\Gamma$ as the five-tuple $(\lambda,\mathcal{T},r,\mathcal{C},u)$. Here, the parameter $\lambda$ corresponds to the mean number of users described by a Poisson random variable with probability mass function defined as
\begin{equation}\label{eq:PoissonPMF}
f(k) = e^{-\lambda} \frac{\lambda^k}{k!}.
\end{equation}
Thus, the number of players in the game is a Poisson random variable with average number of players\footnote{It is important to remark that even though in the original paper of Myerson, the value of $\lambda$ is chosen to be very large, the validity of the results can be applied here as long as a relatively large $\lambda$ is used so that the probability of having zero players in the Poisson game is negligible.} $\lambda >> 1$. Each user from the complete population belongs to one of the types $t \in \mathcal{T}$. The probability of a user to be of type $t$ is given by $r(t)=\text{Prob}(type = t)$. This information is embedded in the vector $r \in \Delta(\mathcal{T})$ where $\Delta(\cdot)$ represents the set of probability distributions over $\mathcal{T}$. By applying the \emph{decomposition property}\cite{BibRef:Myerson98}\cite{BibRef:Sionopoli07} of the Poisson distribution, we can establish that the number of players in the game of type $t$ is also a Poisson random variable with parameter $\lambda r(t)$. On the other hand, it is assumed that the set $\mathcal{C}$ of possible actions is common to all the players regardless of their type. Thus, the set $\Delta(\mathcal{C})$ is the set of mixed actions associated with the players. In Poisson games, the utility of a specific player depends on its type, the action he chooses, and on the number of players (not counting himself), who choose each possible action. The number of players for each possible element in $\mathcal{C}$ is listed in a vector called the \emph{action profile}. Finally, the last term of the tuple is the utility defined as $u = (u_t)_{t\in \mathcal{T}}$ where $u_t(a,x)$ is the payoff that a player of type $t$ receives when a pure action $a$ is chosen and the number of players who choose action $b$ is $x(b)$, for all $b \in \mathcal{C}$.
Now, if the participants play in accordance to the strategy $\sigma$, we call $\sigma_t(a)$ the probability that a player of type $t$ chooses the pure action $a$. Using the decomposition property again, we can establish that the number of players of type $t\in \mathcal{T}$ who choose the pure action $a$ is Poisson distributed with mean $\lambda r(t)\sigma_t(a)$. Since the sum of independent Poisson random variables is also a Poisson variable with mean equal to the sum of the means, the total number of players who take the pure action $a$ is Poisson distributed with mean $\lambda \tau(a)$, where
$$\tau(a) = \sum_{t\in \mathcal{T}} r(t)\sigma_t(a).$$
It follows that, a player of type $t$ who plays a pure action $a \in \mathcal{C}$ while the rest of the players are expected to play using strategy $\sigma$ has a expected utility of
\begin{equation}\label{eq:Expected_Utility_1}
U_t(a,\sigma) = \sum_{x\in Z(\mathcal{C})} P(x|\sigma)u_t(a,x),
\end{equation}
where,
\begin{equation}
P(x|\sigma) = \prod_{b\in \mathcal{C}} e^{-\lambda \tau(b)}\frac{\left(\lambda \tau(b)\right)^{x(b)}}{x(b)!},
\end{equation}
while the expected utility whether the player chooses action $\theta \in \Delta(\mathcal{C})$ is
\begin{equation}\label{eq:Expected_Utility}
U_t(\theta,\sigma) = \sum_{a \in \mathcal{C}}\theta(a)U_t(a,\sigma).
\end{equation}
\subsection{Nash Equilibrium in Poisson Games}
It is very well known that a Nash equilibrium is achieved when each strategy played by all players corresponds to \emph{the best response to all other strategies} in such equilibrium\cite{BibRef:BeibeiWang10}\cite{BibRef:Fudenberg}. Consequently, no player has anything to gain by changing his own strategy unilaterally. The set of best responses for a player of type $t$ against a strategy $\sigma$ is the set of actions that maximizes his expected utility given that the rest of the players (including those whose type is $t$) play as prescribed by $\sigma$. Let us define the set
\begin{equation}\label{eq:NeqRndGames}
  B_t(\sigma) = \left\{b \in \mathcal{C}: b \in \arg \max_{a \in \mathcal{C}} U_t(a,\sigma)\right\}
\end{equation}
as the set of pure best responses against $\sigma$ for a player of type $t$. Equally, the set for mixed best responses against $\sigma$ is the set of actions $B_t(\sigma)=\Delta(B_t(\sigma))$. Therefore, the strategy $\sigma^*$ is a Nash equilibrium if $\sigma^*_t \in B_t(\sigma^*) \forall t$.
\subsection{Examples and Motivation}
\begin{figure}[!t]
\centering
\includegraphics[width = 3.4 in]{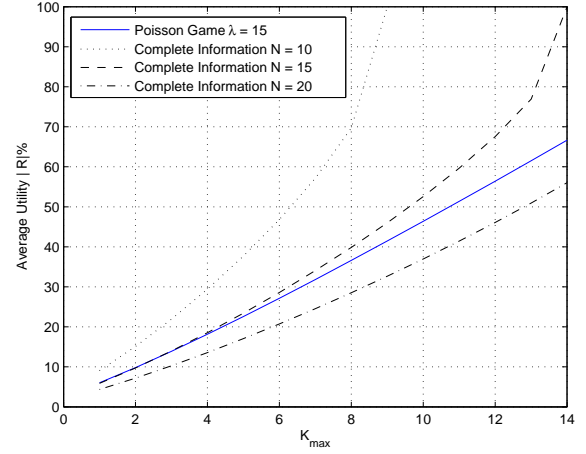}
\caption{Pareto Optimality Utilities for Example 1.}
\label{fig:Utility4}
\end{figure}

\begin{figure}[!t]
\hspace{-15pt}
\centering
\includegraphics[width = 3.8 in]{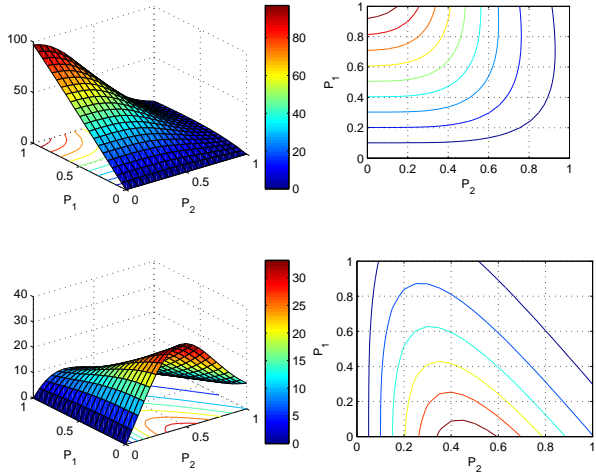}
\caption{Average utility ($\lambda = 15$, $K_{{\max}_1}=7$, $K_{{\max}_2}=5$ and $r_1 = 0.3$).}
\label{fig:Utility7}
\end{figure}

\paragraph{Example 1}
Let $\Gamma$ be a Poisson game with $\lambda = 15$, only one type of players, set of available choices $\mathcal{C} = \left\{ON,OFF \right\}$, and the utility function:
\begin{eqnarray}\nonumber
u(ON,x) &=& \left\{
\begin{array}{cc}
R & \text{if } x(ON)\leq K_{\max} \\
0 & \text{otherwise}
\end{array}\right. , \\
u(OFF,x) &=& 0 \qquad \forall x \in \mathcal{C}, \nonumber
\end{eqnarray}
where $K_{\max}$ is the maximum number of players that can transmit at the same time beside one transmitting player without causing a collision and $R$ is the transmission rate payoff when the player achieves a successful transmission. This game follows the mixed strategies defined as $\sigma(ON) = p$ and $\sigma(OFF) = 1-p$ where $p$ is the probability of transmission by any given player.
Hence, using \eqref{eq:Expected_Utility} we can calculate the expected utility as
\begin{equation}
\hspace{0pt}
\begin{split}
U(p) &= R\sum_{n=1}^{K_{\max}+1}\sum_{i=n}^\infty \frac{n}{i}{i \choose n}p^n(1-p)^{i-n}\left\{e^{-\lambda}\frac{\lambda^i}{i!} \right\}\\ &=
R\sum_{n=1}^{K_{\max}+1} \frac{p^n\lambda^n \left[\Gamma(n+1)-n\Gamma(n,-\lambda(1-p))\right]}{e^\lambda n! (-\lambda(1-p))^n}.
\end{split}
\end{equation}
\begin{figure}[!t]
\centering
\includegraphics[width = 3.4 in]{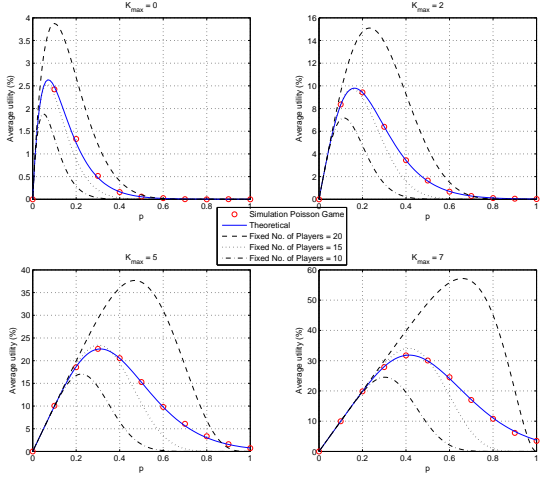}
\caption{Average utility for Poisson Game in Example 1 ($\lambda = 15$).}
\label{fig:Utility1}
\end{figure}
%\begin{figure}[!t]
%\centering
%\includegraphics[width = 3.4 in]{Nash_lambda.eps}
%\caption{Nash equilibrium vs $\lambda$}
%\label{fig:Nash_example1}
%\end{figure}
The dependence of $U(p)$ as a function of $p$ is shown in Figure \ref{fig:Utility1} for different values of $K_{\max}$. The solid line represents the utility of the Poisson game and the dashed lines represent the utility if this was a game with complete information (\emph{i.e.}, fixed number of players known for all). Considering the fact that  a Nash equilibrium predicts in a consistent manner the way in which a game will be played, it is evident that in this game there exists just one logical outcome. In other words, the Nash equilibrium is strict, which by definition must occur in non degenerate strategies\cite{BibRef:Fudenberg}. The equilibrium occurs when all players transmit all the time $(p = 1)$. As noted in Figure \ref{fig:Utility1}, this conveys to a zero utility for all the users. This game is designed in such way that the players have no motivation to not transmit. It can be seen that the utilities obtained are very far from the Pareto optimal\footnote{A Pareto Optimality is defined as a specific set of strategies in which no player can change their strategy and have a greater utility without making any other player utility worse.} which could be achieved by playing a mixed strategy ($p < 1$). Figure \ref{fig:Utility4} shows the achievable utilities if the players were motivated to play Pareto dominant strategies, and thus transmitting only a fraction of time $p < 1$. Notice that by increasing the value of $K_{\max}$ , the probability of transmission by a player increases along side resulting in a higher utility. It is possible to deduce that as $K_{\max}$ tends to the maximum number of players known in a game with complete information, the average utility converges to 100\%. At the same time, in a Poisson game, due to the uncertainty on the number of players, the average utility does not achieve the maximum when $K_{\max}$ tends to $\lambda$. Thus, it is important to consider the Poisson game in detail.
\paragraph{Example 2}
 Consider the Poisson game defined by $\Gamma = \left\{\lambda,\mathcal{T},r,\mathcal{C},u\right\}$, with expected number of players $\lambda = 15$, set of types $\mathcal{T}=\left\{1,2\right\}$ with probabilities $r_1$ and $r_2$, $r_1+r_2=1$ respectively, set of choices $\mathcal{C}=\left\{ON,OFF \right\}$ and the utility function:
% \vspace{-12pt}
\begin{eqnarray}\nonumber
u_1(ON,x) &=& \left\{
\begin{array}{cc}
R_1 & \text{if } x(ON)\leq K_{{\max}_1} \\
0 & \text{otherwise}
\end{array}\right. , \\
u_1(OFF,x) &=& 0 \qquad \forall x \in \mathcal{C}, \nonumber \\
u_2(ON,x) &=& \left\{
\begin{array}{cc}
R_2 & \text{if } x(ON)\leq K_{{\max}_2} \\
0 & \text{otherwise}
\end{array}\right. , \\
u_2(OFF,x) &=& 0 \qquad \forall x \in \mathcal{C}, \nonumber
\end{eqnarray}
where $K_{{\max}_1}$ and $K_{{\max}_2}$ are the maximum number of players who can transmit simultaneously with type 1 and type 2 players respectively. Similarly to Example 1, $R_1$ and $R_2$ are the achievable rates in case of a successful transmission. We define the mixed strategies $\sigma_1(ON) = p_1$ and $\sigma_2(ON) = p_2$ as the transmission probabilities by type 1 and type 2 players respectively. The expected utilities can be calculated by means of eqs. \eqref{eq:Expected_Utility_1}-\eqref{eq:Expected_Utility} as follows
\begin{equation}
\begin{split}
&U_1(p_1,p_2) = R_1 \sum_{n=1}^{K_{{\max}_1}+1}\sum_{k=1}^n \left\{ \sum_{i=k}^{\infty} \frac{k}{i} {i \choose k}p_1^{i}(1-p_1)^{i-k}  \right. \\&  \times \left[e^{-r_1\lambda} \frac{(r_1\lambda)^i}{i!} \right]\sum_{j=n-k}^\infty {j \choose n-k}p_2^j (1-p_2)^{j-n+k} \\ &
\left. \times \left[e^{-r_2\lambda} \frac{(r_2\lambda)^j}{j!} \right]\right\},\\
& U_2(p_1,p_2) = R_2 \sum_{n=1}^{K_{{\max}_2}+1}\sum_{k=1}^n \left\{ \sum_{i=k}^{\infty} \frac{k}{i} {i \choose k}p_2^{i}(1-p_2)^{i-k}  \right. \\ &\times \left[e^{-r_1\lambda} \frac{(r_2\lambda)^i}{i!} \right]\sum_{j=n-k}^\infty {j \choose n-k}p_1^j (1-p_1)^{j-n+k} \\ &
\left. \times \left[e^{-r_1\lambda} \frac{(r_1\lambda)^j}{j!} \right]\right\}.
\end{split}
\end{equation}
Figure \ref{fig:Utility7} shows the expected utility for this example.  Since again the players have no incentive to use mixed strategies, the same strict Nash equilibrium occurs in which all players transmit all the time (\emph{i.e.} $p_1 = p_2 = 1$) resulting in zero utility to all of them. Nevertheless, unlike Example 1, there exist several Pareto dominant mixed strategies in terms of the pairs of probabilities $(p_1,p_2)$. Such a pairs form a Pareto frontier which contains the set of all optimal strategies\cite{BibRef:Fudenberg}\cite{BibRef:Xiao12a}. In the following section, we will reformulate the proposed Poisson games in order to approach to the Pareto optimality.
\section{System Model and Corresponding Game}\label{sec:GameModel}
Let us assume for a moment that there is a fixed and known number $N$ of SUs contending stations while the PU is in the idle state. The transmission queue for all the SUs is assumed to be always nonempty \emph{i.e.} each user always has a packet ready to be transmitted right after the completion of each transmission. We consider that the system works in a slotted ALOHA fashion where each slot of the system is modelled as a one-stage game. At the beginning of each slot, the players have to choose between the two possible actions $\mathcal{C} \in \{ON,OFF\}$ which represent their ability to transmit or to backoff. Every time a player decides to transmit, he can either succeed, in which case he gains throughput, or fail due to a collision , resulting in a penalty (negative throughput) associated with such failure. Moreover, assume that the system is capable of handling multipacket reception (MPR)\cite{BibRef:Chan13} , \emph{i.e.} it is possible to receive several packets simultaneously\footnote{This might be possible by using certain enhancements to the physical layer such as beamforming in MIMO systems, frequency hopping or multiuser detection for instance.}. We consider that the channel has no influence in the loss of any package, therefore the only option for a failure transmission is due to collisions with any package over the MPR limit ($K_{\max} +1$).
\begin{figure}[!t]
\hspace{5pt}
\centering
\includegraphics[width = 3.0 in]{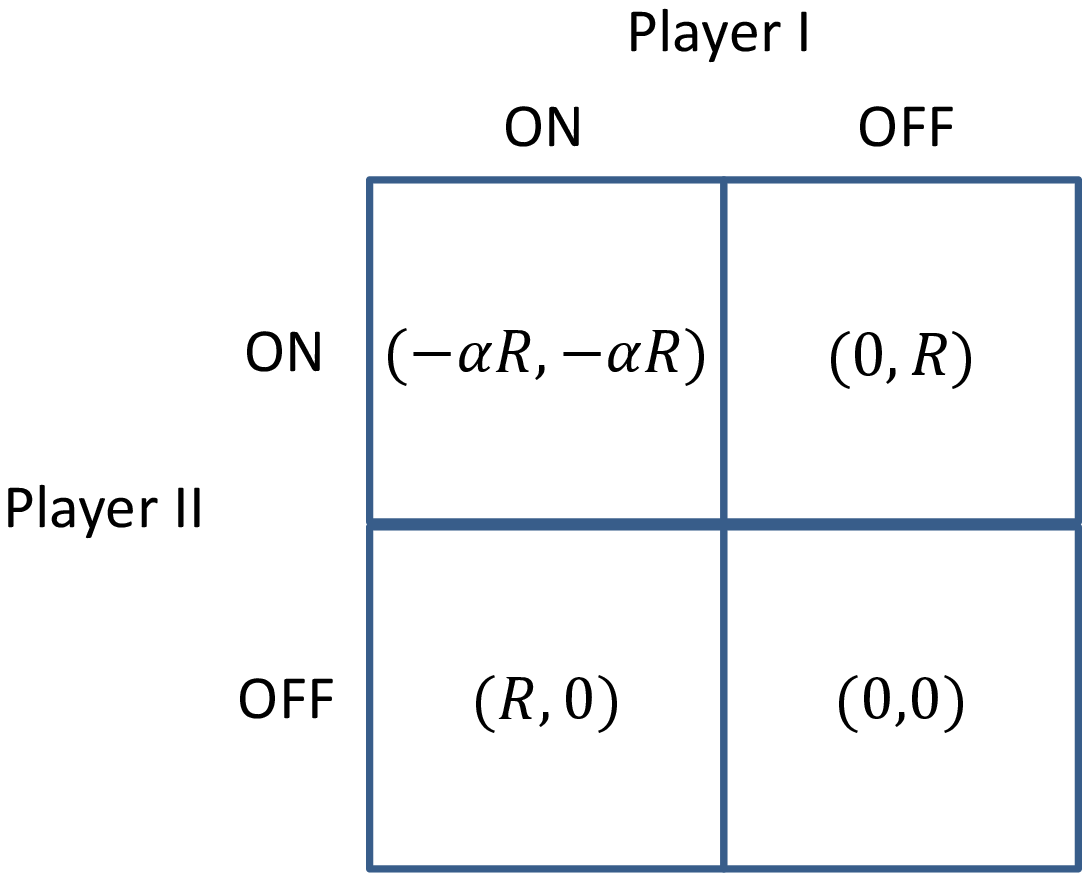}
\caption{Strategic form of multiple access game ($N = 2$).}
\label{fig:Strat_Form}
\end{figure}
As it was shown in Example 1 and Figure \ref{fig:Utility1}, it is possible to obtain the maximum utility by controlling the probability of transmission among the players. This is equivalent to players choosing to play mixed strategies instead of playing the single pure strategy. However, as discussed above, they do not have any incentive to cease transmission to avoid collisions, so, the expected outcome would be all of them transmitting resulting in zero throughput. It is shown in \cite{BibRef:Xiao03} that by introducing a penality to the game it is possible to get closer to the Pareto optimality. Considering the latter, we propose the following game models:
\subsection{Poisson game, single type of players}
Let us reformulate Example 1 by adding a penalty in the case of a collision
\begin{equation}\label{eq:SystemModelFixedNumberOfUsers}
\begin{split}
u(ON,x) &= \left\{
\begin{array}{cc}
R & \text{if } x(ON)\leq K_{\max} \\
-\alpha R & \text{otherwise}
\end{array}\right. , \\
u(OFF,x) &= 0 \qquad \forall x \in \mathcal{C},
\end{split}
\end{equation}
where $\alpha \geq 0$ is a penalization constant. First, we consider a multiple access game with $N \geq 2$ transmitters (players) and $K_{\max} = 1$. In the case of a collision the transmitter is penalized by a constant quantity $-\alpha R$. Figure \ref{fig:Strat_Form} shows the game in strategic form for the case of $N = 2$. An analogous game can be found in \cite{BibRef:Xiao05} as an alternative approach to the distributed coordination function (DCF) in the IEEE 802.11 standard. Each player transmits with probability $p$ following a mixed strategy policy. Consequently, when the number of users is known, the multiple access could be cast as a game with the following utilities:
\begin{equation}\label{eq:Util_fun_mixed_strats}
\begin{split}
\quad U_{OFF,k} &= 0, \\
U_{ON,k} &= p(1-p)^{N-1}R-\alpha R p[1-(1-p)^{N-1}]\\
&=(1-\vartheta)[\vartheta^{N-1}R-\alpha R(1-\vartheta^{N-1})],
\end{split}
\end{equation}
for $k = 1,2,\dots N$. Here we make use of the notation $\vartheta=1-p$. In order for this game to be in equilibrium we need to insure, by choosing a proper penalty $\alpha$, that $U_{OFF,k} = U_{ON,k} = 0, \forall k$. In other words the following should be true
\begin{equation}\nonumber
  \begin{split}
    \vartheta^{N-1}R &=\alpha R(1-\vartheta^{N-1}),\\
    \vartheta &= \left(\frac{\alpha}{1+\alpha} \right)^{\frac{1}{N-1}}.
  \end{split}
\end{equation}
As a result, the mixed strategy in equilibrium given as $p_{eq}$ is achieved by
\begin{equation}
\label{eq:P_eq}
p_{eq} = 1-\left(\frac{\alpha}{1+\alpha} \right)^{\frac{1}{N-1}}.
\end{equation}
It can be seen that
\begin{equation}
p_{eq}=\left\{
\begin{array}{ll}
1 & \text{if } \alpha = 0\\
0 & \text{if } \alpha \rightarrow \infty
\end{array}
\right. .
\end{equation}
Notice that for arbitrary $N \geq 2$ and $\alpha = 0$  (\emph{i.e.} no collision penalty), $p_{eq}=1$.
This shows, as explained in previous section, that in general a game without penalty would not make sense considering that no data can be transmitted at the equilibrium in pure strategies for $N > K_{\max}$.
The amount of data transmitted for a given $\alpha >0$ is then
\begin{equation}
p\vartheta^{N-1}R = \left[1-\left(\frac{\alpha}{1+\alpha}\right)^{\frac{1}{N-1}}\right]\left(\frac{\alpha}{1+\alpha}\right)R.
\end{equation}
The probability of having a particular player transmitting successfully is given as
\begin{equation}
P_{1,k} = p(1-p)^{N-1},
\end{equation}
hence, the maximum of $P_{1,k}$ can be found by taking the partial derivative of its logarithm as follows
$$\frac{\partial \ln P_{1,k}}{\partial p} = \frac{1}{p}-\frac{N-1}{1-p}=0,$$
that results in the unique solution
\begin{equation}
p=\frac{1}{N},
\end{equation}
which is clearly maximum (see also \cite{BibRef:Kang09}). It follows from eq. \eqref{eq:P_eq}, that
$$\frac{1}{N}=1-\left(\frac{\alpha}{1+\alpha}\right)^{\frac{1}{N-1}},$$ and, therefore
\begin{equation}\label{eq:ALPHA_Penalization}
\alpha = \frac{1}{\left(\frac{N}{N-1}\right)^{N-1}-1}.
\end{equation}
If $N = 2$, the corresponding value of $\alpha$ is
$$\alpha(2) = \frac{1}{2-1} = 1.$$On the other extreme, when $N \rightarrow \infty$, one can conclude that
$$\alpha(\infty)=\lim_{N\rightarrow \infty}\frac{1}{\left(\frac{N}{N-1}\right)^{N-1}-1}=\frac{1}{e-1}\approx 0.5$$
Thus, the range of variation of $\alpha$ is $\frac{1}{e-1} \leq \alpha \leq 1$. Since for smaller $N$, collisions are frequent, they require a higher penalty to defer the SUs from continuous transmission.
%\begin{figure}[!t]
%\centering
%\includegraphics[width = 3.5 in]{alpha.eps}
%\caption{Penalization coefficient $\alpha$.}
%\label{fig:alpha}
%\end{figure}
\noindent The case of MPR $K_{max} > 1$ could be treated in a similar fashion as follows. The probability of transmitting without collision can be expressed as
\begin{equation}\label{eq:P_nc_N}
\begin{split}
P_{nc}^{(p)}&=\sum_{k=0}^{K_{\max}} p {N-1 \choose k} p^{k} (1-p)^{N-1-k} \\
&=\sum_{k=0}^{K_{\max}} {N-1 \choose k} p^{k+1}(1-p)^{N-1-k}  \\
&=I_{1-p}(N-1-K_{\max},K_{\max}+1),
\end{split}
\end{equation}
where $I_{1-p}(a,b)$ is the incomplete beta function (see \cite{BibRef:Abramowitz}, chapter 6) and $N>K_{\max}$.
For moderately large $N$, the binomial distribution of interferers could be considered as a sum of $N-1$ binary random variables. Its distribution can be very well approximated by a normal random variable $\xi \sim \mathcal{N}(\mu,\sigma^2)$, where
$$\mu = (N-1)p,$$ $$\sigma^2 = (N-1)p(1-p) = (N-1)pq.$$
Therefore,
\begin{multline}
P_{nc}\approx Prob(\xi < K) = \\ \int_{-\infty}^{K_{max}} \frac{1}{\sqrt{2 \pi (N-1)p q}}\exp\left(\frac{[x-(N-1)p]^2}{2(N-1)pq} \right)dx \\ \approx \Phi \left(\frac{K_{\max}+0.5-(N-1)p}{\sqrt{(N-1)pq}} \right),
\end{multline}
for large $N$ and not so large $K_{\max}$\footnote{One can assume that $P_{nc}(p)$ maximum is achieved when $p << 1$ (or more accurately $p \sim K_{\max}/N$).}
\begin{equation}
K_{\max} < (N-1)p.
\end{equation}
As the next step, let us maximize the last term in the expansion in eq. \eqref{eq:P_nc_N} with respect to $p$,
\begin{equation}
\begin{split}
P_{nc}^{*}(p)&= {N-1 \choose K_{\max}} p^{K_{\max}+1} (1-p)^{N-1-K_{\max}},\\
\frac{\partial \ln P_{nc}^*(p)}{\partial p} &= \frac{K_{\max}+1}{p}-\frac{N-1-K_{\max}}{1-p}=0,
\end{split}
\end{equation}
$$(K_{\max}+1)(1-p)=(N-1-K_{\max})p$$
or
\begin{equation}
p_{\max}=\frac{K_{\max}+1}{N}.
\end{equation}
This term represents the largest contribution to $P_{nc}$ given by eq.$\eqref{eq:P_nc_N}$. Furthermore,
\begin{multline}
\frac{{N-1 \choose K_{\max}} p^{K_{\max}+1} (1-p)^{N-1-K_{\max}}}{{N-1 \choose K_{\max}-1} p^{K_{\max}} (1-p)^{N-K_{\max}}}=\frac{N-K_{\max}}{K_{\max}}\frac{p}{1-p} \\ \approx \frac{Np}{K_{\max}}\frac{1}{1-p}>>1.
\end{multline}
Subsequently, the optimized term provides the bulk contribution to $P_{nc}$. Taking one more term in eq. \eqref{eq:P_nc_N} and following the same reasoning, it is possible to obtain a very accurate approximation to the optimum probability of transmission as
\begin{equation}\label{eq:Approx_Popt}
P_{opt} \approx \frac{K_{\max}+1}{K_{\max}+N}.
\end{equation}
Using this value of $P_{nc}$ we can reformulate eq. \eqref{eq:Util_fun_mixed_strats} as
\begin{equation}\label{eq:UtilityFunctionsSingleType}
\begin{split}
U_{OFF,k} &= 0 \\
U_{ON,k}  &= R P_{nc}(P_{opt})-\alpha R P_{c}(P_{opt})
\end{split}
\end{equation}
where $P_c = 1-P_{nc}$ stands for the probability of collision. Therefore, at the equilibrium
$$U_{OFF,k}=U_{ON,k},$$
$$P_{nc} = \alpha P_c $$
and
\begin{equation}\nonumber
\alpha = \left. \frac{P_{nc}}{1-P_{nc}} \right|_{p=P_{opt}\approx \frac{K_{\max}+1}{K_{\max}+N}}
\end{equation}
\begin{equation}\label{eq:OptAlpha}
 = \left. \frac{\sum_{k = 0}^{K_{\max}}{N-1 \choose k}(1-p)^{N-1-k}p^k}{\sum_{i = K_{\max}}^{N-1}{N-1 \choose i}(1-p)^{N-1-i}p^i} \right|_{p=P_{opt} \approx \frac{K_{\max}+1}{K_{\max}+N}}.
\end{equation}
\begin{figure}[!t]
\centering
\includegraphics[width = 3.5 in]{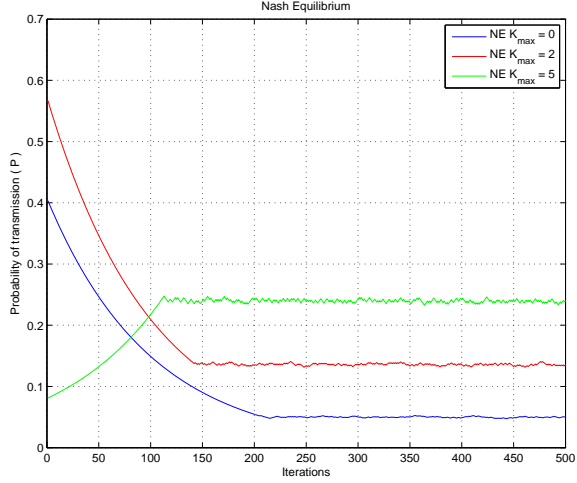}
\caption{Nash equilibrium for different values of $K_{\max}$ and N = 20.}
\label{fig:Nash_Eq_convergence}
\end{figure}
It is consistently assumed in all modelled game theory problems, that the players take decisions based on the most basic notion of rational play where dominated strategies can be iteratively eliminated\cite{BibRef:MackKenzieBook}. It is in this sense that the Nash equilibrium predicts the most likely outcome of the game. The SUs in this game transmit with a certain probability. If such probability coincides with the one in equilibrium, then no player has any incentive to change it. Notwithstanding if the initial probability of transmission is off from the Nash equilibrium, the game described in \eqref{eq:SystemModelFixedNumberOfUsers} will converge to it and will be self-sustaining as can be seen in Figure \ref{fig:Nash_Eq_convergence}.

We consider now that the number of players is a random variable distributed according to a probability distribution $P_N(N)$. Conditioned on the number of players $N$, the pay-off can be rewritten as
\begin{equation}
\begin{split}
\qquad U_{OFF|N} &= 0, \\
U_{ON|N} &= p\vartheta^{N-1}R-p(1-\vartheta^{N-1})\alpha R, \\
&=  p R[(1-\alpha)\vartheta^{N-1}-\alpha].
\end{split}
\end{equation}
Averaging over the distribution of $N$ (eq. \ref{eq:Expected_Utility}), the unconditional utilities can be expressed as
\begin{equation}\label{eq:Unconditional_Payoff}
\begin{split}
U_{OFF} = \sum U_{OFF|N} P_N(N) &= 0 \\
U_{ON} = \sum U_{ON|N} P_N(N) &= \\
 pR\left[(1+\alpha) \sum_{N=1}^{\infty}\vartheta^{N-1}P_N(N)-\alpha \sum_{N=1}^{\infty}P_N(N)\right]& = 0.
\end{split}
\end{equation}
By making use of the following notation
\begin{equation}
\begin{split}
F_N(\vartheta) &= \sum_{N=1}^{\infty}\vartheta^{N-1}P_N(N),
\end{split}
\end{equation}
eq. \eqref{eq:Unconditional_Payoff} can be rewritten as
$$ (1+\alpha) F_N(\vartheta)=\alpha[1-P_N(0)],$$ or
\begin{equation}
\vartheta = F_N^{-1}\left[\frac{\alpha}{1+\alpha}(1-P_N(0))\right],
\end{equation}
where $F_N^{-1}$ is the inverse function of $F_N(\vartheta)$.
For instance, if $P_N(N) = \delta(N-N_0)$ $N_0 > 0$, \emph{i.e.} the game corresponding to a game with complete information (fixed and known number of players), then
\begin{equation}\nonumber
\begin{split}
  P_N(0)&=0, \\
 F_N(\vartheta)&=\vartheta^{N_0-1},
%&\left(\frac{\alpha}{1+\alpha}\right)^{\frac{1}{N_0-1}}
\end{split}
\end{equation}
which coincides with eq. \eqref{eq:P_eq}.
For the Poisson distribution \eqref{eq:PoissonPMF}
\begin{equation}
P_N(N)=\frac{\lambda^N}{N!}e^{-\lambda},
\end{equation}
one can easily obtain
\begin{equation}\nonumber
\begin{split}
P_N(0)&=e^{-\lambda},
\end{split}
\end{equation}
\begin{equation}\label{eq:NumericalInverted}
F_N(\vartheta)=\sum_{N=1}^{\infty} \frac{\vartheta^{N-1}\lambda^N}{N!}e^{-\lambda}=\frac{e^{-\lambda}}{\vartheta}\left[\exp(\vartheta \lambda)-1 \right].
\end{equation}
Notice that eq. \eqref{eq:NumericalInverted} has to be inverted numerically. The required equation for equilibrium is then
\begin{equation}
\begin{split}
\frac{e^{-\lambda}}{\vartheta}\left[\exp(\vartheta \lambda)-1 \right] &= \frac{\alpha}{1+\alpha}\left(1-e^{-\lambda}\right),\\
\frac{e^{\vartheta \lambda}-1}{\vartheta} &= \frac{\alpha}{1+\alpha}\left(e^{\lambda}-1\right).
\end{split}
\end{equation}
If $\alpha = 0$,
$$\frac{e^{\vartheta \lambda}-1}{\vartheta} = 0,$$
does not have solution $\vartheta = 0$ since
$$\lim_{\vartheta \rightarrow 0}\frac{e^{\vartheta \lambda}-1}{\vartheta} = \lambda \neq 0.$$
Consider the reduced Poisson distribution where $N=0$ is not possible. For the case in which there is a random number of transmitters trying to access, the question arises: which penalization $\alpha$ should be chosen in order to obtain the maximum throughput? We can substitute the optimum probability of transmission for each player as
\begin{equation}\label{eq:Popt_approx_random}
\hspace{-10pt}
\begin{split}
P_{opt} \approx \sum_{N=0}^{K_{\max}}\boldmath{1}\cdot \frac{\lambda^N}{N!}e^{-\lambda} + \sum_{N=K_{\max}+1}^\infty \frac{K_{\max}+1}{K_{\max}+N}\frac{\lambda^N}{N!}e^{-\lambda}=\\
\frac{(K_{\max}+1)\left[\Gamma(K_{\max}+1)-K_{\max}\Gamma(K_{\max},-\lambda) \right]}{K_{\max} e^{\lambda}(-\lambda)^{K_{\max}}}.
\end{split}
\end{equation}
Now for the case when $K_{\max} << \lambda$, the first term in  \eqref{eq:Popt_approx_random} can be neglected and using asymptotic of $\Gamma(x,a)$ one obtains the following approximation
\begin{equation}\label{eq:Popt_approx_random2}
P_{opt} \approx \frac{K_{\max}+1}{\lambda+K_{\max}-1}.
\end{equation}
Accordingly, analogously to eq. \eqref{eq:OptAlpha}, we can calculate the optimal $\alpha$ using the following
\begin{equation}\label{eq:OptAlpha_RV}
\hspace{-10 pt}
\alpha = \frac{\sum_{N = K_{\max}+1}^ \infty \sum_{k = 0}^{K_{\max}}{N-1 \choose k}(1-P_{opt})^{N-1-k}P_{opt}^k\frac{\lambda^N}{N!}e^{-\lambda}}{\sum_{N = K_{\max}+1}^ \infty\sum_{i = K_{\max}}^{N-1}{N-1 \choose i}(1-P_{opt})^{N-1-i}P_{opt}^i\frac{\lambda^N}{N!}e^{-\lambda}}.
\end{equation}
\subsection*{Throughput Analysis}
\begin{figure}[!t]
\hspace{-10pt}
\centering
\includegraphics[width = 3.7 in]{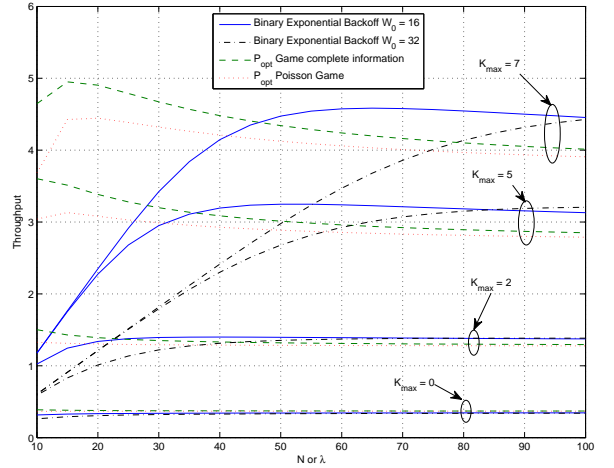}
\caption{Throughput comparison between binary exponential backoff scheme (solid line), game modelled with complete information (dashed line), and game modeled as a Poisson game (dotted line) for different values of $K_{\max}$.}
\label{fig:Throughput_VS_N_Comparison}
\end{figure}

Following the analysis of the MAC with exponential backoff with MPR\cite{BibRef:Kwak05}\cite{BibRef:Zheng06lcn}, we calculate the normalized throughput as follows. First, we obtain the conditional probability of having $k$ packets transmitted successfully given that there was at least one player transmitting in any slot as
\begin{equation}
  P^{(k)}_{succ} = \frac{{N \choose k}P_{opt}^k (1-P_{opt})^{N-k}}{P_{Tx}},
\end{equation}
where $P_{Tx}$ is the probability of having at least one player transmitting in one slot of time which can be computed as
$$P_{Tx} = 1-(1-P_{opt})^N.$$
Therefore, the normalized throughput for the case of a game with complete information is
\begin{equation}\label{eq:T_GPI}
  T = \sum_{k = 1}^{K_{\max}} k P^{(k)}_{succ}  P_{Tx} = \sum_{k = 1}^{K_{\max}} k {N \choose k} P_{opt_N}^k (1-P_{opt_N})^{N-k},
\end{equation}
where $P_{opt_N}$ is given by eq. \eqref{eq:Approx_Popt}. For the case of Poisson game, eq. \eqref{eq:T_GPI} becomes
\begin{multline}
\hspace{15pt}  T_{R} = \sum_{j=0}^{K_{\max}-1}\frac{\lambda^j e^{-\lambda}}{(j-1)!}+ \\
  \sum_{n=K_{\max}}^{\infty}\sum_{k = 1}^{K_{\max}}k {n \choose k} P_{opt_{\lambda}}^k (1-P_{opt_{\lambda}})^{n-k} \frac{\lambda^n e^{-\lambda}}{n!}.
\end{multline}
where $P_{opt_{\lambda}}$ is obtained from eq. \eqref{eq:Popt_approx_random2}. In order to assess the performance of the proposed scheme, we compare the throughput obtained using Poisson games and the one obtained from a binary exponential backoff algorithm implemented in the IEEE 802.11 standard\cite{BibRef:Bianchi00}. Standard contention windows $W_0 = 16$ and $W_0 = 32$ were used for comparison. The results are shown in Figure \ref{fig:Throughput_VS_N_Comparison}. In \cite{BibRef:Kwak05} it is shown that as the number of nodes (players) increases, the throughput converges to a nonzero constant in all cases ($1/2 \ln 2$ for the case of the binary exponential backoff). The case for the game with complete information is also included. Notice that the throughput obtained by means of the Poisson game outperforms the one obtained with the classic exponential backoff for the case of small number of users. This can be explained by the large size of the contention window compared with  the number of users.
\section{A Poisson Game with Two Types of Players }\label{sec:PoissonTwoTypesOfPlayers}
\begin{figure}[!t]
\centering
\includegraphics[width = 3.3 in]{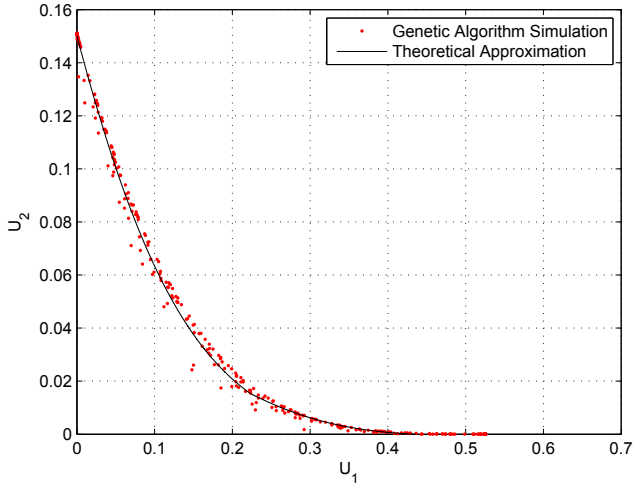}
\caption{Pareto Frontier $(\lambda = 30, r_1 = 0.3, K_{\max_1} = 8, K_{\max_2} = 5)$.}
\label{fig:ProbsP1P2ApproxForPareto_PoissonGames}
\end{figure}

In this section we extend the Poisson game interpretation of Multiple Access to the case of two types of SU, defined by different QoS requirements on their rate. For instance, some users might be using voice services while the rest require a video streaming service. Within this framework we consider that type 1 and type 2 players transmit with rate $R_1$ and $R_2$ respectively. Depending on the QoS for each type of user, we assume that the maximum number of simultaneous transmissions supported by users of first type could be no more than $K_{{\max}_1}+1$ and the maximum supported by the second type could be no more than $K_{{\max}_2}+1$. From here the corresponding Poisson game can be modelled as
\begin{eqnarray}\nonumber
u_1(ON,x) &=& \left\{
\begin{array}{cc}
R_1 & \text{if } x(ON)\leq K_{{\max}_1} \\
-\alpha R_1 & \text{otherwise}
\end{array}\right. , \\
u_1(OFF,x) &=& 0 \qquad \forall x \in \mathcal{C}, \nonumber \\
u_2(ON,x) &=& \left\{
\begin{array}{cc}
R_2 & \text{if } x(ON)\leq K_{{\max}_2} \\
 -\beta R_2 & \text{otherwise}
\end{array}\right. , \\
u_2(OFF,x) &=& 0 \qquad \forall x \in \mathcal{C}, \nonumber
\end{eqnarray}
where $\alpha,\beta >0$ are the two penalization constants in order to guarantee the convergence to a Nash equilibrium in mixed strategies\footnote{Notice that $x(ON)$ is the sum of all transmitting players regardless of their type.}. Similarly to eq. \eqref{eq:UtilityFunctionsSingleType}
for the case of a single type of players, it is possible to write the utilities functions in terms of the probabilities of non collision and collision as
\begin{equation}
\begin{split}
U^{(1)}_{OFF} &= 0, \\
U^{(1)}_{ON}(p_1,p_2)  &= R_1 P^{(1)}_{nc}(p_1,p_2)-\alpha R_1 P^{(1)}_{c}(p_1,p_2),\\
U^{(2)}_{OFF} &= 0, \\
U^{(2)}_{ON}(p_1,p_2)  &= R_2 P^{(2)}_{nc}(p_1,p_2)-\alpha R_2 P^{(2)}_{c}(p_1,p_2).
\end{split}
\end{equation}
 Here $p_1$ and $p_2$ are the probabilities of transmission (mixed strategies) of type 1 and type 2 players respectively. Furthermore, let us just assume, as in the case of single type of players, that this is a game with complete information with $N_1$ and $N_2$ players of each type respectively. The probability of non-collision for both types can be found using the following expressions:
\begin{equation}
\hspace{-25pt}
\begin{split}
&P^{(1)}_{nc}(p_1,p_2) = \sum_{j = 0}^{K_{{\max}_1}}\sum_{i=1}^{j}{N_1 -1 \choose i}{N_2 \choose j-i}p_1^{i+1} (1-p_1)^{N_1-i-1}\\
&\times p_2^{j-i}(1-p_2)^{N_2-j+i},\\
&P^{(2)}_{nc}(p_1,p_2) = \sum_{j = 0}^{K_{{\max}_2}}\sum_{i=1}^{j}{N_2 -1 \choose i}{N_1 \choose j-i}p_2^{i+1} (1-p_2)^{N_2-i-1}\\
&\times p_1^{j-i}(1-p_1)^{N_1-j+i}.
\end{split}
\end{equation}
\begin{figure}[!t]
\centering
\includegraphics[width = 3.8 in]{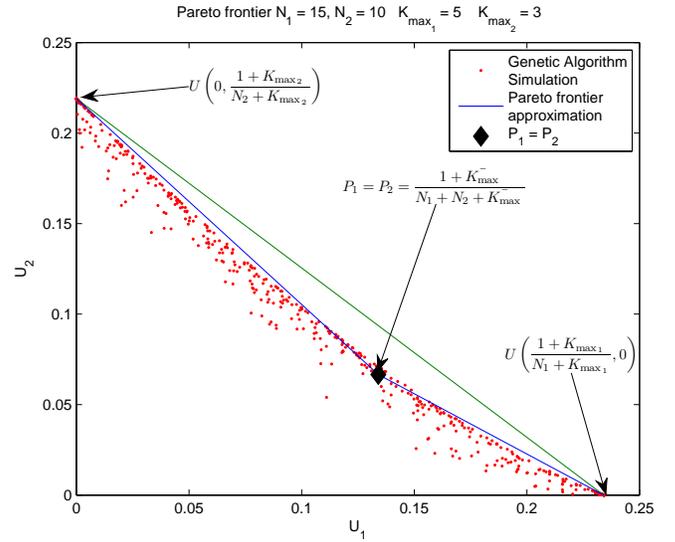}
\caption{Approximation of the Pareto frontier for a complete game with two type of players $(N_1 = 15, N_2 = 10, K_{\max_1} = 5, K_{\max_2} = 3)$.}
\label{fig:Pareto_frontier}
\end{figure}
\noindent As it was shown in Example 2, there is a tradeoff in the choice of $p_1$ and $p_2$. We can see that if $p_1$ is kept fixed and $p_2$ gets decreased, type 1 players are benefited and viceversa. Therefore, it is desirable to assign penalties $\alpha$ and $\beta$ in such a way that the system works in the boundaries of the Pareto frontier, \emph{i.e.} to let all the choices of mixed strategies be Pareto efficient. The question is how to find such a frontier. We start from noticing that if $N_1 = 0$ or $N_2 = 0$ the maximum utility for each type of users is given respectively by eq. \eqref{eq:Approx_Popt} when
$$P^{(1)}_{nc}\left( \frac{1+K_{{\max}_1}}{N_1+K_{{\max}_1}},0 \right) \text{ or } P^{(2)}_{nc}\left(0,\frac{1+K_{{\max}_2}}{N_2+K_{{\max}_2}} \right).$$
In a similar way, using eq. \eqref{eq:Approx_Popt} we can see that for the case of $p_1=p_2$, the maximum probability of non collision for both utilities is achieved by
\begin{equation}
  P = \frac{1+\bar{K}_{\max}}{N_1+N_2+\bar{K}_{\max}},
\end{equation}
where $\bar{K}_{\max}$ is the arithmetic mean of $K_{{\max}_1}$ and $K_{{\max}_2}$. Using these three points, one can construct an approximation to the Pareto frontier by connecting them with straight lines as shown in Figure \ref{fig:Pareto_frontier}. Furthermore, analytical approximation of such frontier as follows.
\begin{equation}\label{eq:P1P2Approx}
\hspace{1pt}p_2 = \left\{
\begin{array}{ll}
m_1 p_1 + \frac{1+K_{\max_2}}{N_2+K_{\max_2}} & 0\leq p_1 \leq \frac{1+\bar{K}_{\max}}{N_1+N_2+\bar{K}_{\max}} \\
m_2\left(p_1 - \frac{1+K_{\max_1}}{N_1+K_{\max_1}} \right) & \frac{1+\bar{K}_{\max}}{N_1+N_2+\bar{K}_{\max}} \leq p_1 \leq \frac{1+K_{\max_1}}{N_1+K_{\max_1}}
\end{array}
\right.
\end{equation}
where $m_1$ and $m_2$ are calculated respectively as
\begin{equation}\label{eq:P1P2Approx_2}
\begin{split}
m_1 &= \frac{\left(\frac{1+\bar{K}_{\max}}{N_1+N_2+\bar{K}_{\max}}-\frac{1+K_{\max_2}}{N_2+K_{\max_2}}\right)\left(N_1+N_2+\bar{K}_{\max} \right)}{1+\bar{K}_{\max}},\\
m_2 &= -\frac{1+\bar{K}_{\max}}{\left(N_1+N_2+\bar{K}_{\max} \right)\left(\frac{1+K_{\max_1}}{N_1+K_{\max_1}}-\frac{1+\bar{K}_{\max}}{N_1+N_2+\bar{K}_{\max}}\right)}.
\end{split}
\end{equation}
The accuracy of such approximation achieved by eq. \eqref{eq:P1P2Approx} is shown in Figure \ref{fig:Pareto_frontier}. We compare the approximation calculated with the Pareto frontier obtained by a simulation using a genetic algorithm with 30 iterations in Matlab\cite{BibRef:Mitchell}. All solutions in a Pareto set are equally optimal, so it is up to the wireless designer to select a solution in that set depending on the application or the QoS goal. Furthermore, extending this to the case of a Poisson game, the probability of non collision for both types can be obtained using the following equations
\begin{equation}\label{eq:PoissonGameModel2types}
\begin{split}
 P^{(1)}_{nc} = \sum_{j=0}^{K_{{\max}_1}}\sum_{i=0}^{j}\mathcal{P}_1(i) \mathcal{P}_2(j-i), \\
 P^{(2)}_{nc} = \sum_{j=0}^{K_{{\max}_2}}\sum_{i=0}^{j}\mathcal{P}_1(j-i) \mathcal{P}_2(i),
\end{split}
\end{equation}
where
\begin{equation}
\hspace{-15 pt}
  \mathcal{P}_n(x) = \sum_{s = 0}^{x-1}\frac{(r_n \lambda)^{s}}{s!}e^{-r_n \lambda}+\sum_{l=x}^\infty
  {l \choose x} p_n^x(1-p_n)^{l-x} \frac{{(r_n \lambda)}^l}{l!} e^{-r_n \lambda}.
\end{equation}
The Pareto frontier for the case of Poisson Games can be calculated as
\begin{equation}\label{eq:P1P2Approx_RandomGames}
\hspace{-18pt}p_2 = \left\{
\begin{array}{ll}
m_1 p_1 + \frac{1+K_{\max_2}}{r_2 \lambda + K_{\max_2}-1} & 0\leq p_1 \leq \frac{1+\bar{K}_{\max}}{\lambda+\bar{K}_{\max}} \\
m_2\left(p_1 - \frac{1+K_{\max_1}}{r_1 \lambda +K_{\max_1} -1} \right) & \frac{1+\bar{K}_{\max}}{\lambda + \bar{K}_{\max}} \leq p_1 \leq \frac{1+K_{\max_1}}{r_1 \lambda +K_{\max_1} -1},
\end{array}
\right.
\end{equation}
where $m_1$ and $m_2$ are calculated respectively as
\begin{equation}
\begin{split}
m_1 &= \frac{\left(\frac{1+\bar{K}_{\max}}{\lambda+\bar{K}_{\max}}-\frac{1+K_{\max_2}}{r_2 \lambda +K_{\max_2}-1}\right)\left(\lambda +\bar{K}_{\max} \right)}{1+\bar{K}_{\max}},\\
m_2 &= -\frac{1+\bar{K}_{\max}}{\left(\lambda+\bar{K}_{\max} \right)\left(\frac{1+K_{\max_1}}{r_1 \lambda +K_{\max_1}-1}-\frac{1+\bar{K}_{\max}}{\lambda + \bar{K}_{\max}}\right)},
\end{split}
\end{equation}
following the same technique as in deriving eqs. \eqref{eq:P1P2Approx} and \eqref{eq:P1P2Approx_2}. The Pareto Frontier for the Poisson game with two types of players is shown in Figure \ref{fig:ProbsP1P2ApproxForPareto_PoissonGames}. The solid line represents the utility obtained by using eq. \eqref{eq:P1P2Approx_RandomGames} to calculate the mixed strategies in eq. \eqref{eq:PoissonGameModel2types}.
Consequently, the necessary condition for the system to be in equilibrium is
\begin{equation}
\begin{split}
U^{(1)}_{OFF} &= U^{(1)}_{ON}(p_1,p_2) = 0 \\
U^{(2)}_{OFF} &= U^{(2)}_{ON}(p_1,p_2) = 0
\end{split}
\end{equation}
thus the penalization factor $\alpha$ and $\beta$ can be respectively obtained as
\begin{equation}\label{eq:OptPenalTwoTypes}
  \alpha = \frac{P^{(1)}_{nc}(p_1,p_2)}{1-P^{(1)}_{nc}(p_1,p_2)} \text{ and } \beta = \frac{P^{(2)}_{nc}(p_1,p_2)}{1-P^{(2)}_{nc}(p_1,p_2)}.
\end{equation}
Here $p_1$ and $p_2$ are obtained from the Pareto frontier by means of eq. \eqref{eq:P1P2Approx_RandomGames}.

\section{Game Model considering Primary User Activity}\label{sec:GameModelwithPUactivity}
In this section we consider the existence of a single PU transmitting within the same channel as the SUs. We assume that the PU transmits in a slot by slot basis with probability $P_{T}$ and that the slots are synchronized between the PU and the SUs. In this sense, we consider a PU transmitting to be in an ON state. Let $\bar{N}_{ON}$ be the average number of consecutive slots in which the PU is in an ON state. Furthermore, let us observe the PU over $\nu >> \bar{N}_{ON}$ sequential time slots. Then, on average, there will be $\nu P_{T}$ ON states and therefore the average number of transitions from OFF to ON state is simply
\begin{equation}
\eta_{{}_{OFF\rightarrow ON}}\approx \frac{ \nu P_{T}}{\bar{N}_{ON}}.
\end{equation}
We assume that all the SUs have perfect detection of the PU activity, hence when the latter is transmitting, all SUs remain silent. Let $P_{SU,T}$ be the probability that there is a transmi\-ssion from one or more SU when the state of the PU is turned ON for the first time after being in an OFF state and by consequence, creating a collision with the PU. The average number of collisions $\bar{N}_{col}$ is given by
\begin{equation}
\bar{N}_{col} = P_{SU,T} \cdot  \eta_{{}_{OFF\rightarrow ON}} = \frac{\nu P_{T}P_{SU,T} }{\bar{N}_{ON}}.
\end{equation}
Consequently, the average probability of collision between SUs and the PU is
\begin{equation}\label{eq:PUavgColliProb}
\begin{split}
P_{col,PU} = \frac{\bar{N}_{col} }{\nu P_{T}} =\frac{P_{SU,T} }{\bar{N}_{ON}}.
\end{split}
\end{equation}
For the game with full information, the aforementioned probability $P_{SU,T}$ represents the probability of having at least one SU transmitting at the same time as the PU. This can be calculated as follows
\begin{figure}[!t]
  \centering
  \includegraphics[width=3.5in]{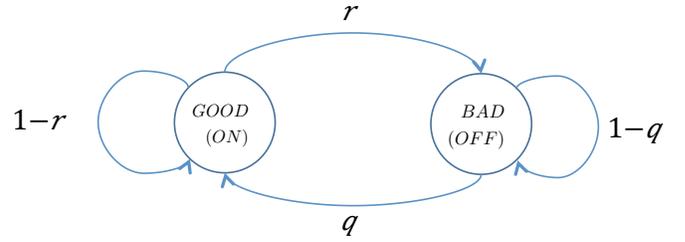}
  \caption{Primary User Activity}\label{fig:PUActiviy}
\end{figure}
\begin{equation}\label{eq:PSUTSingePlayer}
P_{SU,T}(p) = 1-\left(1-p\right)^{N}.
\end{equation}
As an example of the previous, let us assume that the transmissions from a single PU operate in a channel inversion with cut-off mode\cite{BibRef:Goldsmith97}. The transmissions follow a Gilbert-Elliot (GE) model (particularly one which imposes a correlation $\rho$ in the time domain) \cite{BibRef:Kanal78}\cite{BibRef:Primak12iccit} where the ``GOOD'' state occurs with probability $P_{T}$ and the ``BAD'' state occurs with probability $1-P_{T}$. In this case, the probability of transmission $P_{T}$ is directly related to the fading channel as 
\begin{equation}
P_{T} = \int_{\gamma_{0}}^{\infty} p_{\gamma}(\gamma) d\gamma
\end{equation}
where $\gamma_{0}$ is an energy threshold above which a transmission is possible and $p_{\gamma}(\gamma)$ is the fading distribution of the channel. Within this context, we consider that the PU will transmit in a slot only within a ``GOOD'' state and remain silent on the other case. As it can be seen in Figure \ref{fig:PUActiviy}, we can model the dynamic traffic from the PU with $r$ and $q$ defined as follows\cite{BibRef:Primak12iccit}
\begin{equation}
\begin{split}
q &= P_{T}(1-\rho), \\
 r &= (1-P_{T})(1-\rho),
\end{split}
\end{equation}
where $0 \leq \rho \leq 1$ is the correlation coefficient.  The average duration of the PU in the ON state $\bar{N}_{ON}$ can be calculated as 
\begin{equation}
\bar{N}_{ON}=\sum_{i=1}^{\infty} i (1-r)^{i-1} r = \frac{1}{r} = \frac{1}{(1-P_{T})(1-\rho)}.
\end{equation}
Let  $P_{col}^{Th}$ be a pre-established tolerance threshold defined as the maximum average probability of collision $P_{col,PU} $  the PU could be able to tolerate. It follows from equation \eqref{eq:PUavgColliProb} that
\begin{equation}\label{eq:ConditionProtectPU}
P_{SU,T} (p)\leq \bar{N}_{ON}  P_{col}^{Th}.
\end{equation}
Notice that for $P_{col}^{Th} \geq 1/ \bar{N}_{ON}$ any value of $p$ satisfies eq. \eqref{eq:ConditionProtectPU}. This means that the SUs can transmit with any probability and just stop transmitting when they detect the presence of a transmitting PU. On the other hand, when $P_{col}^{Th}  < 1/ \bar{N}_{ON}$,  there is a value $p^{*}$ such that
\begin{equation}\label{eq:Popt4PU}
P_{SU,T} (p^{*}) = \bar{N}_{ON} P_{col}^{Th},
\end{equation}
which can be rewritten as
\begin{equation}\label{eq:PstarN}
p^*(N) =  1-\left(1-\bar{N}_{ON} P_{col}^{Th}\right)^{\frac{1}{N}}. 
\end{equation}
If $p^{*} \geq \frac{K_{\max}+1}{K_{\max}+N}$ in eq. \eqref{eq:Approx_Popt}, the impact to PU can be once again ignored. Nevertheless, if $p^{*} < \frac{K_{\max}+1}{K_{\max}+N},$ a different value of $P_{opt}$ needs to be used in order to calculate the penalty $\alpha$ in eq. \eqref{eq:OptAlpha}. 
\begin{figure}[!t]
  \centering
  \includegraphics[width=3.5in]{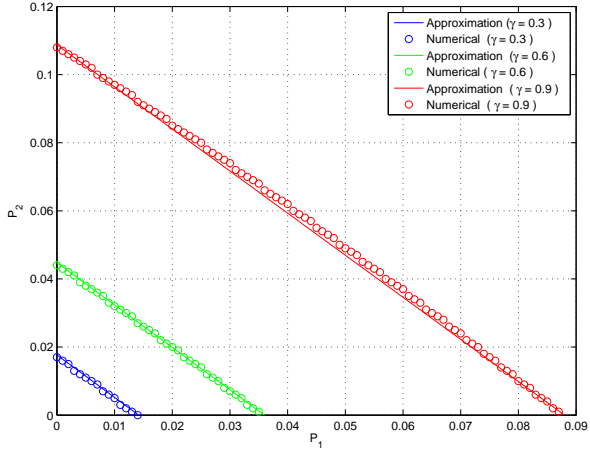}
  \caption{Restriction frontier approximation for $N_{1} = 25, N_{2} = 20$.}\label{fig:RestrictionFrontierFullInfo}
\end{figure}
Consequently $P_{opt}$ in eq. \eqref{eq:Approx_Popt} will be given as
\begin{equation}\label{eq:UltimatePopt}
P_{opt} = \min\left(p^*(N), \frac{K_{\max}+1}{K_{\max}+N}\right).
\end{equation}
Considering the analyzed example, one can notice that for a fixed tolerance $P_{col}^{Th}$, the SUs can transmit using the optimal strategy without significantly affecting the PU if the latter has very poor channel conditions and/or the channel is uncorrelated.
By extending this analysis to the case of two types of players, the analogous effect of eq. \eqref{eq:UltimatePopt} is to create a restriction frontier under which the SUs can calculate their penalization factor $\alpha$ and $\beta$ by means of equation \eqref{eq:OptPenalTwoTypes}. Such a frontier can be obtained by rewriting eq. \eqref{eq:PSUTSingePlayer} as
\begin{equation}\label{eq:PSUTTwoPlayer}
P_{SU,T} (p_{1},p_{2}) = 1-(1-p_{1})^{N_{1}}(1-p_{2})^{N_{2}} \leq \bar{N}_{ON}  P_{col}^{Th},
\end{equation}
and noticing that the maximum $P_{SU,T}$ for each type of player that satisfies eq. \eqref{eq:PSUTTwoPlayer} occurs when either all type 1 SUs transmit with probability $p^{*}_{1}(N_{1})$ and none of the type 2 SUs transmit (\emph{i.e.} $P^{(1)}_{SU,T} \left(p^{*}_{1}(N_{1}),0 \right)$), or, all type 2 SUs transmit with probability $p^{*}_{2}(N_{2})$ and none of the type 1 SUs transmit (\emph{i.e.}  $P^{(2)}_{SU,T} \left(0,p^{*}_{2}(N_{2}) \right))$.
Hence, an accurate approximation of the restriction frontier can be formed by connecting the aforementioned two points with a straight line as shown in Figure \ref{fig:RestrictionFrontierFullInfo}. It can be seen from the previous that the equality of such an approximation is very good for the selected values of parameters. We omit a detailed analysis due to lack of space.
Finally, in order to calculate $p^{*}$ for a Poisson game with the average number of SU denoted as $\lambda$, ones has to average eq. \eqref{eq:PstarN} over the Poisson distribution  as
\begin{equation}\label{eq:pPUactPoisson}
\begin{split}
p^{*}(\lambda)&=\sum_{k=0}^{\infty} p^{*}(k) \frac{\lambda^{k}e^{-\lambda}}{k!} \\ &=\sum_{k=0}^{\infty} \left\{1-\left(1-\bar{N}_{ON}  P_{col}^{Th} \right)^{\frac{1}{k}} \right\}\frac{\lambda^{k}e^{-\lambda}}{k!}.
\end{split}
\end{equation}
Here $p^{*}(k)$ can be expanded in terms of a Taylor series with respect to the number of players $k$ around the average value $\lambda$ to produce
\begin{equation}
\begin{split}
p^{*}(k) &= e^{\frac{\ln (1-\bar{N}_{ON}  P_{col}^{Th})}{\lambda}}-e^{\frac{\ln (1-\gamma)}{\lambda}}\frac{\ln(1-\bar{N}_{ON}  P_{col}^{Th})(k-\lambda)}{\lambda^{2}}  \\ &+ \mathcal{O}\left\{(k-\lambda)^{2} \right\}.
\end{split}
\end{equation}
By taking the first two terms of the latter, we can approximate the solution of eq. \eqref{eq:pPUactPoisson} for large $\lambda$ (as assumed throughout this paper) as
\begin{equation}
p^{*}(\lambda) \approx 1- \frac{\lambda e^{\lambda} - \lambda - \ln \left(1-\bar{N}_{ON}  P_{col}^{Th} \right)}{\lambda e^{\lambda - \ln \left(1-\bar{N}_{ON}  P_{col}^{Th} \right)}}.
\end{equation}
Therefore $P_{opt}$ in eq. \eqref{eq:Popt_approx_random2} can be obtained simply by
\begin{equation}
P_{opt} = \min\left(p^{*}(\lambda),\frac{K_{\max}+1}{K_{\max}+\lambda-1} \right).
\end{equation}
\begin{figure}[!t]
  \centering
  \includegraphics[width=3.1in]{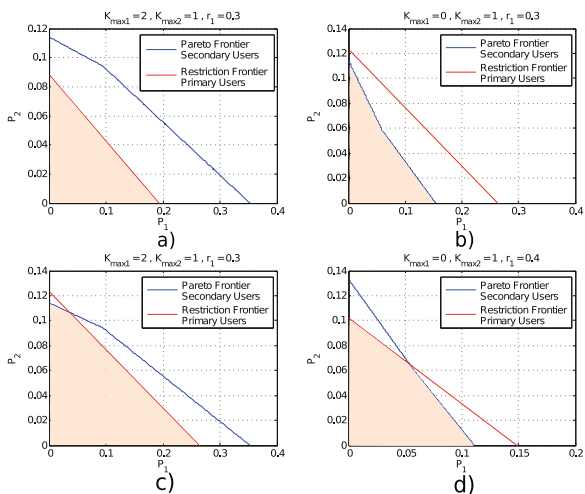}
  \vspace{-10 pt}
  \caption{Restriction Frontier and ParetoFrontier for Poisson Games   $(\lambda = 25, \bar{N}_{ON}  P_{col}^{Th} = 0.9).$}\label{fig:RestrictionFrontierVsParetoFrontierPoisson}
\end{figure}
Analogously, for the case of two types of players eq. \eqref{eq:PSUTTwoPlayer} becomes
\begin{equation}
\begin{split}
P_{SU,T} (p_{1},p_{2}) = &\sum_{i=0}^{\infty}\sum_{j=0}^{\infty}\left\{ 1-(1-p_{1})^{i}(1-p_{2})^{j} \right\} \times \\ e^{\lambda} \frac{(r_{1}\lambda)^{i}}{i!}\frac{(r_{2}\lambda)^{j}}{j!} & \leq \bar{N}_{ON}  P_{col}^{Th}.
\end{split}
\end{equation}
Thus the restriction frontier can be formed, as seen in the previous case simply by connecting the two points corresponding to 
$$P^{(1)}_{SU,T} \left(p^{*}_{1}(r_{1}\lambda),0 \right) \text{ and } P^{(2)}_{SU,T} \left(0,p^{*}_{2}(r_{2}\lambda) \right).$$
Figure \ref{fig:RestrictionFrontierVsParetoFrontierPoisson} shows the influence of the PU activity on the strategy choice by SUs in the case of two type of players for different values of $K_{\max}$. It is possible to distinguish three different scenarios. In Figure  \ref{fig:RestrictionFrontierVsParetoFrontierPoisson}(a) one can see that the restriction frontier lies below the Pareto frontier. This means that the SUs have to limit their transmission probabilities within the boundaries of the shown area in order to not significantly affect the PU performance. On the contrary, in Figure \ref{fig:RestrictionFrontierVsParetoFrontierPoisson}(b), the Pareto frontier is found to be below the restriction frontier. Here, the SUs can choose their strategies based on the Pareto frontier already calculated with the guarantee that PU's performance will remain unaltered and at the same time, the maximum throughput will be obtained. In Figures \ref{fig:RestrictionFrontierVsParetoFrontierPoisson}(c)-(d), we encounter the case where there is an intersection between the Pareto frontier and the restriction frontier. Here it can be observed that the SUs can use any transmission strategy within the area shown in order to keep the harm provoked to the PU to the minimum as in previous cases. However, it is always better to choose strategies which lie on the Pareto frontier boundaries in contrast to the restriction frontier boundaries in order to obtain a better throughput to all players. 

\section{Conclusion}\label{sec:Conclusion}
This paper presents a game-theoretic perspective on the secondary users multiple access problem in cognitive radio networks. In the first part of the paper we have shown how to design a fixed number of homogeneous SUs game with penalization. The Nash equilibrium from such a game results in an optimal throughput of the network per secondary user. Corresponding analytical expressions have been obtained for the SU nodes with MPR properties. Furthermore, we have extended these results to be incorporated into a game with a random number of players (SUs). The Poisson Games allows us to account for a dynamically changing situation on the number of active SUs.  An optimal probability of transmission for the SUs (a mixed strategy) is calculated in order to achieve the Pareto optimality in the system. The latter was proved to achieve a better performance for small number of users than other well known approaches such as the DCF. The next part of the paper extends the game-theoretic analysis to the case of two different types of SUs with different QoS requirements. We showed that the Pareto frontier can be accurately approximated by means of connecting a piece-wise function based on the optimal probability of transmission obtained from each type of players in isolation. Finally, we have considered  the impact of the dynamic activity of the PU on the SU optimal strategy. It is shown that the optimal probabilities of transmission for the SUs are influenced by the PU pattern only under certain conditions. More specifically, we have derived the conditions for the Pareto frontier under which the SU activities are limited by either the equilibrium strategy in  games without a PU, by constrains on the SINR of a PU, or by both in a piece-wise manner. In particular, it was shown that if the PU has low intermittency in the transmitting intervals, its affects on the SU strategy could be neglected. However, for PUs with relatively bursty activities the strategy of SUs is limited by the SINR requirements at the PU.

\bibliography{IEEEabrv,References}

\begin{thebibliography}{10}
\providecommand{\url}[1]{#1}
\csname url@samestyle\endcsname
\providecommand{\newblock}{\relax}
\providecommand{\bibinfo}[2]{#2}
\providecommand{\BIBentrySTDinterwordspacing}{\spaceskip=0pt\relax}
\providecommand{\BIBentryALTinterwordstretchfactor}{4}
\providecommand{\BIBentryALTinterwordspacing}{\spaceskip=\fontdimen2\font plus
\BIBentryALTinterwordstretchfactor\fontdimen3\font minus
  \fontdimen4\font\relax}
\providecommand{\BIBforeignlanguage}[2]{{%
\expandafter\ifx\csname l@#1\endcsname\relax
\typeout{** WARNING: IEEEtran.bst: No hyphenation pattern has been}%
\typeout{** loaded for the language `#1'. Using the pattern for}%
\typeout{** the default language instead.}%
\else
\language=\csname l@#1\endcsname
\fi
#2}}
\providecommand{\BIBdecl}{\relax}
\BIBdecl

\bibitem{BibRef:MackKenzieBook}
A.~{MacKenzie} and L.~{DaSilva}, \emph{Game Theory for Wireless
  Engineers}.\hskip 1em plus 0.5em minus 0.4em\relax USA: Morgan \& Claypool
  Publishers, 2006.

\bibitem{BibRef:ZhangBook}
\BIBentryALTinterwordspacing
Y.~Zhang and M.~Guizani, \emph{Game Theory for Wireless Communications and
  Networking}, ser. Wireless Networks and Mobile Communications.\hskip 1em plus
  0.5em minus 0.4em\relax Taylor and Francis, 2011. [Online]. Available:
  \url{http://books.google.ca/books?id=YKWtPwAACAAJ}
\BIBentrySTDinterwordspacing

\bibitem{BibRef:BeibeiWang10}
\BIBentryALTinterwordspacing
B.~Wang, Y.~Wu, and K.~R. Liu, ``Game theory for cognitive radio networks: An
  overview,'' \emph{Comput. Netw.}, vol.~54, no.~14, pp. 2537--2561, Oct. 2010.
  [Online]. Available: \url{http://dx.doi.org/10.1016/j.comnet.2010.04.004}
\BIBentrySTDinterwordspacing

\bibitem{BibRef:Feng13}
\BIBentryALTinterwordspacing
X.~Feng, H.~Wang, and X.~Wang, ``A game approach for cooperative spectrum
  sharing in cognitive radio networks,'' \emph{Wireless Communications and
  Mobile Computing}, pp. n/a--n/a, 2013. [Online]. Available:
  \url{http://dx.doi.org/10.1002/wcm.2364}
\BIBentrySTDinterwordspacing

\bibitem{bibRef:Gharehshiran13}
O.~Gharehshiran, A.~Attar, and V.~Krishnamurthy, ``Collaborative sub-channel
  allocation in cognitive lte femto-cells: A cooperative game-theoretic
  approach,'' \emph{Communications, IEEE Transactions on}, vol.~61, no.~1, pp.
  325--334, 2013.

\bibitem{BibRef:Zhang12}
Y.~Zhang, D.~Niyato, P.~Wang, and E.~Hossain, ``Auction-based resource
  allocation in cognitive radio systems,'' \emph{Communications Magazine,
  IEEE}, vol.~50, no.~11, pp. 108--120, 2012.

\bibitem{BibRef:Kasbekar12}
G.~Kasbekar and S.~Sarkar, ``Spectrum pricing games with random valuations of
  secondary users,'' \emph{Selected Areas in Communications, IEEE Journal on},
  vol.~30, no.~11, pp. 2262--2273, 2012.

\bibitem{BibRef:MacKenzie03infocom}
A.~MacKenzie and S.~Wicker, ``Stability of multipacket slotted aloha with
  selfish users and perfect information,'' in \emph{INFOCOM 2003. Twenty-Second
  Annual Joint Conference of the IEEE Computer and Communications. IEEE
  Societies}, vol.~3, march-3 april 2003, pp. 1583 -- 1590 vol.3.

\bibitem{BibRef:Xiao05}
Y.~Xiao, X.~Shan, and Y.~Ren, ``Game theory models for {IEEE} 802.11 {DCF} in
  wireless ad hoc networks,'' \emph{Communications Magazine, IEEE}, vol.~43,
  no.~3, pp. S22 -- S26, march 2005.

\bibitem{BibRef:Bianchi00}
G.~{Bianchi}, ``Performance analysis of the {IEEE 802.11} distributed
  coordination function,'' \emph{{IEEE} J. Sel. Areas Commun.}, vol.~18, no.~3,
  pp. 535 --547, March 2000.

\bibitem{BibRef:DandanWang06}
\BIBentryALTinterwordspacing
D.~Wang, C.~Comaniciu, and U.~Tureli, ``Cooperation and fairness for slotted
  aloha,'' \emph{Wirel. Pers. Commun.}, vol.~43, no.~1, pp. 13--27, Oct. 2007.
  [Online]. Available: \url{http://dx.doi.org/10.1007/s11277-006-9240-5}
\BIBentrySTDinterwordspacing

\bibitem{BibRef:Bae13}
Y.~H. Bae, ``Analysis of optimal random access for broadcasting with deadline
  in cognitive radio networks,'' \emph{Communications Letters, IEEE}, vol.~17,
  no.~3, pp. 573--575, 2013.

\bibitem{BibRef:Park11}
J.~{Park} and M.~{Van Der Schaar}, ``Cognitive {MAC} protocols using memory for
  distributed spectrum sharing under limited spectrum sensing,'' \emph{{IEEE}
  Trans. Commun.}, vol.~59, no.~9, pp. 2627--2637, 2011.

\bibitem{BibRef:Cali00}
F.~Cali, M.~Conti, and E.~Gregori, ``Dynamic tuning of the ieee 802.11 protocol
  to achieve a theoretical throughput limit,'' \emph{Networking, IEEE/ACM
  Transactions on}, vol.~8, no.~6, pp. 785 --799, dec 2000.

\bibitem{BibRef:Kwak05}
B.-J. Kwak, N.-O. Song, and L.~Miller, ``Performance analysis of exponential
  backoff,'' \emph{Networking, IEEE/ACM Transactions on}, vol.~13, no.~2, pp.
  343 -- 355, april 2005.

\bibitem{BibRef:Zheng06lcn}
P.~X. Zheng, Y.~J. Zhang, and S.~C. Liew, ``Analysis of exponential backoff
  with multipacket reception in wireless networks,'' in \emph{Local Computer
  Networks, Proceedings 2006 31st IEEE Conference on}, nov. 2006, pp. 855
  --862.

\bibitem{BibRef:Ghez88}
S.~{Ghez}, S.~{Verdu}, and S.~{Schwartz}, ``Stability properties of slotted
  aloha with multipacket reception capability,'' \emph{{IEEE} Trans. Autom.
  Control}, vol.~33, no.~7, pp. 640 --649, July 1988.

\bibitem{BibRef:Choi03icc}
S.~Choi, J.~del Prado, S.~S. N, and S.~Mangold, ``{IEEE} 802.11 e
  contention-based channel access ({EDCF}) performance evaluation,'' in
  \emph{Communications, 2003. ICC '03. IEEE International Conference on},
  vol.~2, may 2003, pp. 1151 --1156 vol.2.

\bibitem{BibRef:Milchtaich03}
\BIBentryALTinterwordspacing
I.~Milchtaich, ``Random-player games,'' \emph{Games and Economic Behavior},
  vol.~47, no.~2, pp. 353--388, May 2004. [Online]. Available:
  \url{http://ideas.repec.org/a/eee/gamebe/v47y2004i2p353-388.html}
\BIBentrySTDinterwordspacing

\bibitem{BibRef:Myerson98}
\BIBentryALTinterwordspacing
R.~B. Myerson, ``Population uncertainty and {P}oisson games,''
  \emph{International Journal of Game Theory}, vol.~27, no.~3, pp. 375--392,
  1998. [Online]. Available:
  \url{http://ideas.repec.org/a/spr/jogath/v27y1998i3p375-392.html}
\BIBentrySTDinterwordspacing

\bibitem{BibRef:Myerson00}
\BIBentryALTinterwordspacing
------, ``{L}arge {P}oisson {G}ames,'' \emph{Journal of Economic Theory},
  vol.~94, no.~1, pp. 7--45, September 2000. [Online]. Available:
  \url{http://ideas.repec.org/a/eee/jetheo/v94y2000i1p7-45.html}
\BIBentrySTDinterwordspacing

\bibitem{BibRef:Koskie05}
S.~Koskie and Z.~Gajic, ``A nash game algorithm for sir-based power control in
  3g wireless cdma networks,'' \emph{Networking, IEEE/ACM Transactions on},
  vol.~13, no.~5, pp. 1017 -- 1026, oct. 2005.

\bibitem{BibRef:Sionopoli07}
\BIBentryALTinterwordspacing
F.~D. Sionopoli and C.~G. Pimienta, ``Undominated (and) perfect equilibria in
  {P}oisson games,'' Universidad Carlos III, Departamento de Economía,
  Economics Working Papers we073117, Apr. 2007. [Online]. Available:
  \url{http://ideas.repec.org/p/cte/werepe/we073117.html}
\BIBentrySTDinterwordspacing

\bibitem{BibRef:Fudenberg}
D.~Fudenberg and J.~Tirole, \emph{Game Theory}.\hskip 1em plus 0.5em minus
  0.4em\relax Cambridge, MA: MIT Press, 1991.

\bibitem{BibRef:Xiao12a}
Y.~Xiao, J.~Park, and M.~Van~der Schaar, ``Repeated games with intervention:
  Theory and applications in communications,'' \emph{Communications, IEEE
  Transactions on}, vol.~60, no.~10, pp. 3123--3132, 2012.

\bibitem{BibRef:Chan13}
D.~S. {Chan}, T.~{Berger}, and L.~{Tong}, ``Carrier sense multiple access
  communications on multipacket reception channels: Theory and applications to
  {IEEE} 802.11 wireless networks,'' \emph{{IEEE} Trans. Commun.}, vol.~61,
  no.~1, pp. 266 --278, January 2013.

\bibitem{BibRef:Xiao03}
\BIBentryALTinterwordspacing
M.~Xiao, N.~B. Shroff, and E.~K.~P. Chong, ``A utility-based power-control
  scheme in wireless cellular systems,'' \emph{IEEE/ACM Trans. Netw.}, vol.~11,
  no.~2, pp. 210--221, Apr. 2003. [Online]. Available:
  \url{http://dx.doi.org/10.1109/TNET.2003.810314}
\BIBentrySTDinterwordspacing

\bibitem{BibRef:Kang09}
L.~Kang and L.~Ni, ``Revisiting {ALOHA} with physical layer network coding,''
  technical report, CSE Dep. HKUST, Tech. Rep., 2009.

\bibitem{BibRef:Abramowitz}
M.~Abramowitz and I.~Stegun, Eds., \emph{Handbook of Mathematical
  Functions}.\hskip 1em plus 0.5em minus 0.4em\relax New York: Dover, 1965.

\bibitem{BibRef:Mitchell}
M.~Mitchell, \emph{An Introduction to Genetic Algorithms}.\hskip 1em plus 0.5em
  minus 0.4em\relax Cambridge, MA, USA: MIT Press, 1998.

\bibitem{BibRef:Goldsmith97}
A.~{Goldsmith} and P.~{Varaiya}, ``Capacity of fading channels with channel
  side information,'' \emph{{IEEE} Trans. Inf. Theory}, vol.~43, no.~6, pp.
  1986--1992, November 1997.

\bibitem{BibRef:Kanal78}
L.N.Kanal and A.R.K.Sastry, ``Models for channels with memory and their
  applications to error control,'' \emph{Proceedings of the IEEE}, vol.~66,
  no.~7, pp. 724--744, July 1978.

\bibitem{BibRef:Primak12iccit}
S.~{Primak}, ``On a dynamic model of cognitive radio systems,'' in
  \emph{Communications and Information Technology (ICCIT), 2012 International
  Conference on}, June 2012, pp. 144 --148.

\end{thebibliography}
\bibliographystyle{IEEEtran}
\end{document}